\documentclass[%
 reprint,
 amsmath,amssymb,
 aps,
prl,
]{revtex4-1}

\usepackage{graphicx}
\usepackage{dcolumn}
\usepackage{bm}

\usepackage{amssymb} 
\usepackage{braket}
\usepackage{mathcomp}
\usepackage{amsmath}

\begin{document}

\preprint{APS/123-QED}

\title{Quantum entanglement and the non-orientability of spacetime}

\author{Ovidiu Racorean}
 \email{ovidiu.racorean@mfinante.gov.ro}
\affiliation{%
 General Direction of Information Technology, Bucharest, Romania\\
}%

\date{\today}

\begin{abstract}
We argue, in the context of Ads/CFT correspondence, that the degree of entanglement on the CFTs side determines the orientation of space and time on the dual global spacetime. That is, the global spacetime dual to entangled copies of field theory is non-orientable, while the product state of the CFTs results in an orientable spacetime. As a result, disentangling the degrees of freedom between two copies of CFT implies, on the gravity side, the transition from a non-orientable spacetime to a spacetime having a definite orientation of space and time, thus an orientable spacetime. We conclude showing that topology change induced by decreasing the entanglement between two sets of degrees of freedom corresponds to a topological blow down operation. 

\end{abstract}

                            
\maketitle


\section{\label{sec:level1}Introduction 
}
Quantum entanglement is one of the most counter-intuitive properties of quantum systems. In recent years, the concept of entanglement has been applied into the understanding about the high energy and gravity physics. This unveiled some remarkable connections between quantum information theory and gravity. In the context of AdS/CFT duality, the structure of quantum entanglement in the CFTs state is intimately related to the geometrical structure of the dual spacetime \cite{mal}, \cite{sus}.  Furthermore, it has been argued in \cite{mal}, \cite{sus}, \cite{raa}, \cite{mark}, that the emergence of classically connected spacetimes is the consequence of the entanglement structure of the underlying quantum mechanical degrees of freedom. The initial conditions of the correlations between the CFTs influence the dual geometry of spacetime. Therefore, the initial product state of the CFTs (no initial correlation between the two quantum subsystems) is dual to two disconnected AdS spacetimes, while classical connectivity between the spacetime pair arises as result of initial entangled states.

In this Letter we will argue that if this scenario is correct, than the spacetime connecting the two regions of quantum system should be non-orientable. Accordingly, we will first probe the time non-orientability of dual spacetime by considering another interesting insight into the phenomenology of quantum entanglement that comes from emergence of the thermodynamic arrow of time. Thus, the same argument of initial conditions between two quantum systems has been reiterated in recent debate \cite{par}, \cite{seth}, \cite{jen}, \cite{jev}, \cite{mac}, \cite{parr}, \cite{mic} on the emergence of thermodynamic arrow of time. Exciting progresses have been reported considering the argument of initial conditions between two quantum systems as principal factor in determining the orientation of the arrow of time. Therefore, it has been shown \cite{mic}, based on the changing of entropy, that the lack of initial correlations between the two quantum systems results in the emergence of a preferred direction of the arrow of time, as is reflected by the standard second law. On the other hand, initial high-correlation environment allow both orientations of the thermodynamic arrow.

We explore further the arguments of initial conditions of the emergence of thermodynamic arrow of time in the framework of AdS/CFT duality \cite{rac}. We argue, that the orientation of the thermodynamic arrow on the dual asymptotically AdS spacetime is determined by the degree of entanglement between the two copies of CFTs. 

Second, the arguments on the space non-orientability of the dual spacetime are constructed around the space and time reversal symmetry of the thermofield double state \cite{isr} ,\cite{goh}, \cite{cap}, \cite{cot}, \cite{ber}. Consequently, we advocate that the time non-orientability is accompanied by the space non-orientability of the global spacetime. 

Based on the initial conditions, we conclude that the structure of the global spacetime geometry is consistent with that of a non-orientable spacetime, in the presence of the initial high-correlations, while in the case of low initial correlations the geometry is that of an oriented spacetime. 

We examine the topology change of global spacetime by expanding on van Raamsdonk idea \cite{raa} of disentangling the degrees of freedom between the two copies of field theory. Traditionally, it has been argued, that disentangling the degrees of freedom between CFTs to zero, the shortest bulk path  \cite{bal1},  \cite{bal2},  \cite{bal3},  \cite{bal4},  \cite{vil} between the corresponding regions in the field should gradually tend to infinity, that is the regions are disconnected from each other. We have a smooth topology change as the two regions gradually pull apart in direct proportionality to the decrease in entanglement. 

We further approach the above scenario from the perspective of a global non-orientable spacetime as dual to entangled CFTs. Thus, on the gravity side, this scenario is consistent with the following statement: starting with a global non-orientable spacetime and decreasing to zero the entanglement of the dual copies of CFT, the resulting spacetime is orientable.  Roughly speaking, disentangling the degrees of freedom between two copies of field theory implies, on the geometry side, a transition from a non-orientable spacetime to a spacetime having a definite orientation of space and time, thus an orientable spacetime. Therefore, starting with initial high-correlation and disentangling the degrees of freedom is consistent, in the geometrical dual, to a topological blow down operation \cite{sha}, \cite{kol}, \cite{man}, \cite{sta}. Since disentangling the degrees of freedom between regions is equivalent to a topological blow down, we conclude that the change in topology is an abrupt transition from a non-orientable spacetime to an orientable spacetime.

\section{\label{sec:level2}Initially uncorrelated states}

We start considering two non-interacting copies of CFT on sphere $S^d$. The two CFTs are equivalent to two subsystems noted left ($L$) and right ($R$), such that we can decompose Hilbert space $\mathcal{H}_{LR}$ of the composite system as  $\mathcal{H}_{LR}=\mathcal{H}_L\otimes \mathcal{H}_R$ . We consider that the two subsystems, $L$ and $R$, respectively are initially uncorrelated to begin with. Thus, in this case, since there is no entanglement between the two CFTs components, we have two completely separate physical systems which do not interact, such that the joint state of the system is the product state: 

 \begin{equation}
\rho_{LR}=\rho_L\otimes \rho_R.
\end{equation}. 				                              
Here, quantities, $\rho_L$, $\rho_R$ and $\rho_{LR}$ are the density matrix describing the states of the left, right ant the composite system, respectively. It is important to note here that $\rho_L$ and $\rho_R$ correspond to the thermal state of the left and right copies of field theory. 

From this initial product state the correlations between the two quantum systems can only increase. To motivate this statement and measure the correlations between the two subsystems, $L$ and $R$, we consider here the mutual information:

\begin{equation}
I(\rho_{LR} )= S(\rho_L )+S(\rho_R )-S(\rho_{LR} ),
\end{equation}

where $S(\rho_L )$, $S(\rho_R )$ and $S(\rho_{LR})$ are the entropies of the left, right and the composite system, respectively. We have considered here, as usual, the von Neumann entropy, $S(\rho)=-Tr(\rho log\rho)$, of the density matrix , to define the entropies.  

Since we are in a low-entropy environment, due to the product state in Eq. (1), we know that , $I(\rho_{LR})= 0$, which reduces the Eq. (2) to: 

\begin{equation}
S(\rho_{LR})=S(\rho_L)+S(\rho_R).
\end{equation}
                                       
As a result, the mutual information and consequently the entropy of the composite system can only increase. 
Let us now the two systems interact in isolation from the rest of the universe such that, gradually, some of the degrees of freedom of the individual components become entangled, evolving in this way the entropy of the joint state from the initial, $S_i(\rho_{LR})=S_i(\rho_L)+S_i(\rho_R)$ ,	 to the final entropy $S_f(\rho_{LR})\leq S_f(\rho_L)+S_f(\rho_R)$.

The initial and final states of the composite system, are related unitarily, thus $S_f (\rho_{LR})=S_i (\rho_{LR})$, such that we have:

\begin{equation}
\Delta S(\rho_L )+\Delta S(\rho_R )\geq0.
\end{equation}

This result is consistent with the second law of thermodynamics which states that the entropy of an isolated system can only increase. Generally, the total entropy can only increase since the mutual information is zero. In this case, the absence of initial correlations between the two CFT’s, the entropy can only increase, in accordance with the formulation of the second law of thermodynamics, such that we have a precise direction of the arrow of time oriented in the sense of increasing entropy. Specifically, the flow of time is directed toward the standard thermodynamic arrow.

On the gravity side of the AdS/CFT correspondence, the interpretation of this initially uncorrelated state is straightforward \cite{raa}. The two separate physical quantum systems determined by the density matrix $\rho_L$ and $\rho_R$, correspond in the dual description, to a spacetime with two disconnected asymptotically AdS regions. Taking into account the emergence of the thermodynamic arrow of time, we may conclude that on the gravity side, the structure of the geometric dual is that of a spacetime with a preferred orientation of time. We can argue here, that the lack of initial correlations between CFTs ensures a time-orientation on the dual spacetime.

\section{\label{sec:level3}Initially entangled states}

Let us now consider that the two copies of the field theory are high-correlated to begin with. In this scenario the situation changes dramatically. To be more specific, we further consider that the two copies of CFT are initially entangled in the thermofield double state. Thus, the joint state of the two subsystems can be represented, in the context of AdS/CFT correspondence, as:

\begin{equation}
\rho_{LR}=\ket{\Psi_\beta} \bra{\Psi_\beta},
\end{equation}

with $\ket{\Psi_\beta}$   defined as the thermofield double state,

\begin{equation}
\ket{\Psi_{\beta}} = \frac{1}{\sqrt{Z_{\beta}}}\sum_ne^\frac{-\beta E_n}{2}{\ket{n}}_L{\ket{n}}_R,
\end{equation}

where   $\ket{n}_L (\ket{n}_R)$ is the $n$-th energy eigenvector for the subsystem $L (R)$ , $\beta$ is the inverse temperature and ${Z_\beta}^{-\frac{1}{2}}$ is a normalization constant. Since Eq.(6) is essentially the Schmidt decomposition of $\ket{\Psi_\beta}$  , we may remark here that the two subsystems are initially entangled in a pure state. Note however that the individual states of the two so constructed subsystems are the thermal states. We can see this by tracing over the Hilbert space of the other, $\rho_L={Tr}_R \rho_{LR}$ (and $\rho_R={Tr}_L \rho_{LR}$).  

It has been argued in  \cite{mal},\cite{sus},\cite{raa},\cite{mark} that the thermofield double state $\ket{\Psi_\beta}$   corresponds, on the gravity side, to the eternal AdS black hole spacetime, whose extended Penrose diagram is depicted in Fig.(1). 

\begin{figure}
\includegraphics[width=8.6cm]{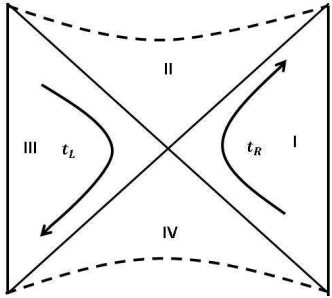}
\caption{\label{fig:fig1} Extended Penrose diagram of AdS eternal black hole. Note that the time coordinate t runs in opposite directions in the two asymptotic regions associated with the quadrants I and III.}
\end{figure}

We further expand on the Van Raamsdonk’s idea \cite{raa} on disentangling some of the degrees of freedom between the two entangled CFTs and ask what happens with the entropy of the composite system once we decrease the degree of entanglement of the composite system. 

Since we are now in a high-correlation environment it was shown in \cite{raa} that disentangling some of the degrees of freedom is consistent with the decrease of mutual information. Consequently, the entropy also decreases. 

In stark contrast to initially uncorrelated case, we can emphasize here that the initial pure state of the two copies of the field theory ensures that $\rho_L$ and $\rho_R$ are isospectral so that $S(\rho_L )=S(\rho_R )$. Let us now start from the thermofield double state and gradually disentangle the degrees of freedom. The initial entanglement of the composite system forces the individual entropies $S(\rho_L)$ and $S(\rho_R)$ to move in the same direction, such that, $\Delta S(\rho_L)=\Delta S(\rho_R )$, at all times. In addition, we can say that the initial pure state, implies that the individual entropies can only decrease, such that  $\Delta S(\rho_L)=\Delta S(\rho_R)<0$. 

In this case, Eq.(4) is reversed as:

\begin{equation}
\Delta S(\rho_L )+\Delta S(\rho_R )\leq 0,
\end{equation}

a result which suggests that ,from the perspective of the second law of thermodynamics, both orientations of the thermodynamic arrow are allowed, such that there is no opportunity for the dominance of one direction of time over the other \cite{par}, \cite{seth}, \cite{jen}, \cite{jev}, \cite{mac},\cite{parr}, \cite{mic}.  

On the gravity side, it has been argued in \cite{mal}, \cite{sus}, \cite{raa} that the geometric structure of the dual spacetime of the entangled CFTs is that of a spacetime with two asymptotically AdS regions classically connected. Moreover, considering the changes in the entropy discussed above, from the perspective of AdS/CFT correspondence, the dual spacetime of the composite system is a spacetime with no preferred orientation of time. The initial high correlations of the two CFT’s ensure that in the dual spacetime the arrow of time may be oriented normal, from the past to the future or in reverse, from the future to the past. In this context, since there is no preferred orientation of time on the spacetime, we may conclude that the gravity dual is a time-unoriented spacetime with two classically connected regions. 

\section{\label{sec:level4}The structure of global spacetime}

We have seen that the global spacetime dual to entangled CFTs should be a time non-orientable manifold. We would like to ask, at this point, what is the space structure of entangled CFTs dual global specetime. In particular we would like to know whether the global spacetime is also space non-orientable.

Intuitively it seems clear that time non-orientability of spacetime should trigger also the space non-orientability. This statement is based on the space reversal (parity) and time reversal symmetry acting on the thermofield double state. That is, the logic tells us that in order to preserve the symmetry the dual spacetime should be space non-orientable. 
To see this more clearly, recall that earlier discussions \cite{isr} have been pointed out that thermofield dynamics encodes a reflexive symmetry between the left and right halves of an eternal black hole depicted in the extended Penrose diagram in Fig.(1). Quantum systems $L$ and $R$ correspond to field modes propagating in the left and right quadrants, while the thermally entangled state $\ket{\Psi_{\beta}}$ corresponds to the ground state on the global spacetime manifold.

From this perspective, we start on the CFT side, with the two copies of field theory entangled in the thermofield double state. One simple diagnostic of entanglement as a non-vanishing connected two-point function at space-like separation. Considering operators $\mathcal{O}_L$ acting on  ${CFT}_L$ and operators $\mathcal{O}_R$ acting on  ${CFT}_R$ (in points $x_L$ and $x_R$), we can write the two-sided correlator (connected two-point function of operators) by the analytical continuation of $t$ as \cite{goh}, \cite{cap}:

\begin{widetext}
\begin{equation}
\begin{split}
\bra{\Psi_{\beta}} \mathcal{O}_L (x_L,0)\mathcal{O}_R (x_R,t)\ket{\Psi_{\beta}}
& =\sum_{n,m} e^{-\frac{\beta}{2} (E_n+E_m) + it(E_n-E_m)}\bra{n}_L \mathcal{O}_L (x_L,0) \ket{m}_L \bra{n}_R \mathcal{O}_R (x_R,0) \ket{m}_R \\
& =\sum_{n,m} e^{-{\beta} E_n + i(t-i \frac{\beta}{2})(E_n-E_m)}\bra{n}_L \mathcal{O}_L (x_L,0) \ket{m}_L \bra{m}_R {\mathcal{O}_R}^\dagger (x_R,0) \ket{n}_R
\end{split}
\end{equation}
\end{widetext}

We can express the correlators in the TFD state as the correlators on the cylinder, that is, if the operators $\mathcal{O}_L$ in ${CFT}_L$ are located at  $\tau=0$ , the operators $\mathcal{O}_R$ are located at  $\tau=\frac{\beta}{2}$ , at the opposite side on the cylinder. 

Since the correlators are situated in opposite sides of the cylinder, it seems legitimate to assume that the continuation in Eq. (9) is $t \rightarrow t - i\pi$ , as argued in \cite{goh}, \cite{cap}. Thus, the thermofield double continuation takes quadrant I of the extended Penrose diagram to quadrant III.  More precise, on the extended Penrose diagram we take two copies of the exterior region and two boundaries corresponding to the eternal $(AdS)_(d+1)$ -Schwarzschild black hole with the metric: 

\begin{equation}
{ds}^2=-f(r) {dt}^2 + {f(r)}^{-1} {dr}^2 + r^2 d{{\Omega}^2}_{d-1}
\end{equation}

where $f(r)=r^2+1-c_d GMr^{(2-d)}$, and $c_d=8(d-1)^{-1} \pi^{[\frac{(2-d)}{2}]} \Gamma(\frac{d}{2})$. 
We introduce the tortoise coordinate $\frac{(dr_*)}{dr}= {f(r)}^{-1}$ such that the right boundary is at $r_*= 0$ and the future horizon is at  $r_* \rightarrow -\infty$, $t \rightarrow \infty$. With the help of Kruskal variables,  $U=-e^{(\frac{2\pi}{\beta})(r_*-t)}$ and $V=e^{\frac{(2\pi)}{\beta})(r_*+t)}$, the metric is smoothly extended past the horizon in quadrant I. The future horizon is then at $U =0$ , with V finite. The past horizon is at $V=0$, with U finite, that is, in region III, connected to the left asymptotic region, we have $U=e^{\frac{(2\pi)}{\beta})(r_*-t)}$ , $V=-e^{\frac{(2\pi)}{\beta})(r_*+t)}$.

If we consider the continuation,  $t \rightarrow t - i\pi$, in the Kruskal variables on the right halve of Penrose diagram we see that  $(U,V) \rightarrow (-U,-V)$, that is, the thermofield continuation takes quadrant I to quadrant III. Then $(U,V) \rightarrow (-U,-V)$ is symmetry of the AdS metric, so to speak, the thermofield double encodes space and time reversal symmetry. Since we have seen the spacetime dual to the two copies of field theory is time non-orientable, it implies, to preserve the space and time reversal symmetry, that it also is space non-orientable. 

Since the thermofield double state describe the eternal black hole than the global spacetime must have a space and time non-orientable structure. As such, we conclude that the full spacetime dual to two entangled copies of field theory is global non-orientable.

\section{\label{sec:level5}Topology change}

We would like to expand on the Van Raamsdonk’s idea \cite{raa} which states that decreasing the entanglement between the degrees of freedom of two copies of field theory lead to topology change in the dual spacetime and see what happens in the case of global non-orentable spacetime. 

Let us advocate a scenario in which we start with the two copies of the field theory initially entangled in the thermofield double state and gradually disentangle the degrees of freedom to zero. Since, initially, the CFTs are entangled, the mutual information, $I(\rho_{LR} )$, has a nonzero value proportional to the number of degrees of freedom in the gauge theory lying on the boundary of AdS. Traditionally, using Ryu-Takayanagi proposal \cite{ryu}, it has been argued, that decreasing the entanglement is equivalent, in the dual spacetime, with a decrease of the minimal surface which separates the two regions of the spacetime, $L$ and $R$, the two CFTs are living on. Accordingly, the two regions of dual spacetime, initially connected, gradually pinch off from each other in direct proportionality to the decrease of entanglement. In this case we have $S(\rho_{LR} )=S(\rho_L )+S(\rho_R )$ and a sharp vanishing of $I(\rho_{LR} )$,  then occurs. When the mutual information vanishes, the reduced density matrix $\rho_{LR}$ factorizes into $\rho_{LR}=\rho_L \otimes \rho_R$, implying that the two regions are completely decoupled from each other and thus, all the correlations (both classical and quantum) between regions $L$ and $R$ should be rigorously zero.

To see this more clearly, recall that mutual information is related to correlations between operators $\mathcal{O}_L$ and $\mathcal{O}_R$ acting on subsystems $L$ and $R$ through, \cite{raa}, \cite{wol}:

\begin{equation}
I(L,R) \geq \frac{(< \mathcal{O}_L(x_L) \mathcal{O}_R (x_R) > - < \mathcal{O}_L (x_L) > < \mathcal{O}_R (x_R) >)^2}{2 {\| \mathcal{O}_L(x_L) \|}^2 {\| \mathcal{O}_R (x_R) \|}^2}
\end{equation}

When all correlations between the two subsystems decrease, the distance in the bulk between the two regions increase \cite{bal1}, \cite{bal2}, \cite{bal3}, \cite{bal4}, \cite{vil} according to:

\begin{equation}
< \mathcal{O}_L (x_L) \mathcal{O}_R(x_R) > \sim e^{-m \mathcal{L}_{bulk}(x_L,x_R)}
\end{equation}

where $\mathcal{L}$ is the length of the shortest path in the bulk that relates $x_L$ and $x_R$.

For high correlations, as is the case of maximally entangled CFTs, the distance is minimal and tends to zero. As the correlations between degrees of freedom in region $L$ and region $R$ decrease to zero the length of path between the regions expand to infinity.

We will see how the above scenario changes from the perspective of non-orientable spacetime. We have argued that the spacetime dual to entangled state of two copies of field theory is a non-orientable manifold, while the product state of two CFTs is dual to an orientable spacetime. As a result, the structure of the spacetime manifold ${M_e}^d$ dual to entangled CFTs should include the real projective plane, ${\mathbb{R} P}^d$ . Consequently, the product state of two CFTs is dual to an orientable spacetime manifold $M^d$.

Having this in mind, we start with the two entangled copies of CFT having a dual spacetime manifold ${M_e}^d$ that contains ${\mathbb{R} P}^d$ and decrease to zero the entanglement between the degrees of freedom in regions $L$ and $R$. From the gravity dual perspective, the above scenario is translated in the following statement: starting with a global non-orientable spacetime manifold ${M_e}^d$ and decreaseing to zero the entanglement of the dual copies of CFT, the resulting spacetime is the orientable manifold, $M^d$. Roughly speaking, disentangling the degrees of freedom between two copies of field theory implies, on the geometry side, a transition from the non-orientable spacetime to a spacetime having a definite orientation of space and time, thus an orientable spacetime. The topology of spacetime changes from a non-orientable manifold to an orientable manifold. Based on this argument we note here that disentangling the degrees of freedom between the two CFTs is equivalent to topologically blow down procedure of spacetime manifold ${M_e}^d$ to $M^d$.

We argue here that the initial spacetime is the connected sum of the orientable spacetime $M^d$ dual to the pair of CFTs in product state with the real projective plane, ${\mathbb{R} P}^d$:

\begin{equation}
{M_e}^d=M^d \# {\mathbb{R} P}^d
\end{equation}

More precisely, the spacetime dual to entangled pair of field theory is the topological blow up of the orientable spacetime dual to CFTs in product state.

To build intuition of what happens to the dual geometry in the non-orientable spacetime scenario let us exemplify with the simplest case of CFTs on a sphere,  $S^2$. Now, we take two subsystems corresponding to regions L and R on the sphere as is depicted in the right panel of Fig.(2). The CFT is a local quantum field theory such that we can consider specific degrees of freedom associated to the regions $L$ and $R$ as maximally entangled. The correlations between the two regions have the maximum value such that, in this case, according to Eq. (12), the length of the shortest path ($\mathcal{L}$) in the bulk between the points $x_L$ and $x_R$ of these regions tend to zero, $\mathcal{L} \rightarrow 0$. In addition, the maximal entanglement between the degrees of freedom associated with the two regions results in the global non-orientable spacetime with the simplest ${\mathbb{R} P}^2$ topology.

\begin{figure}
\includegraphics[width=8.6cm]{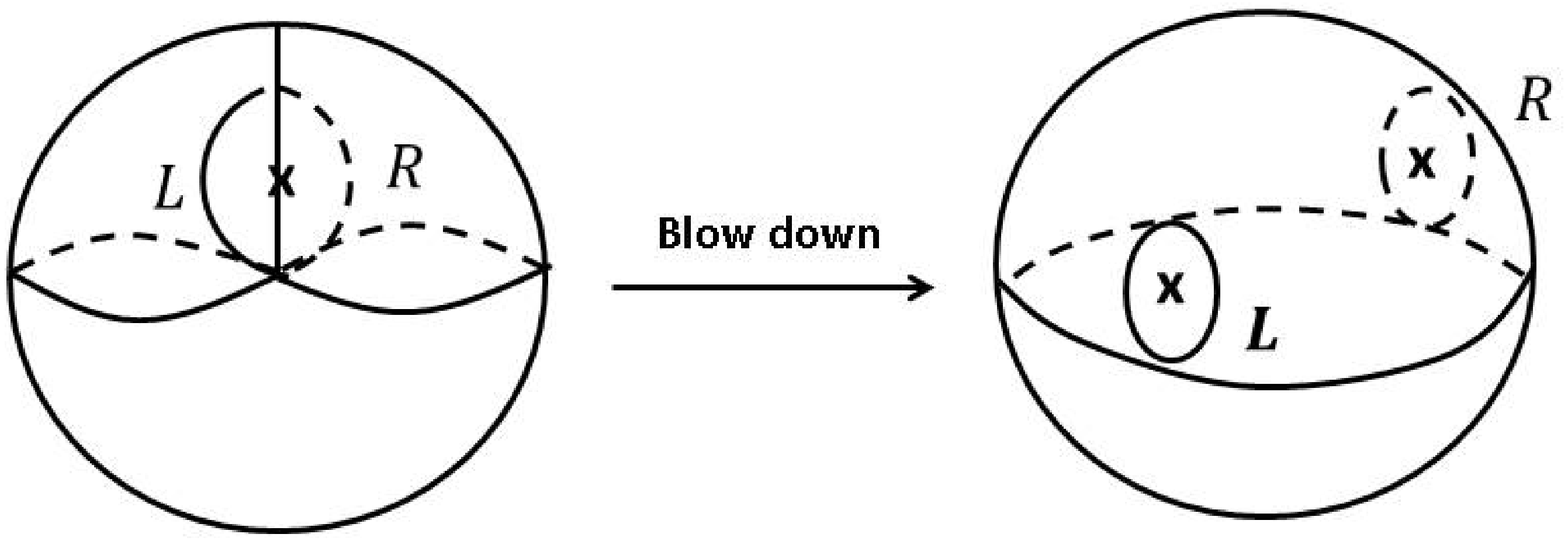}
\caption{\label{fig:fig2} Connection of regions L and R corresponding to entangled copies of field theory form a non-orientable spacetime (pictured here as cross-cap). The dashed lines are on the back side. Disentangling the degrees of freedom associated to regions L an R corresponds to a topological blow down operation.}
\end{figure}

Since the blow up topology of a sphere is equivalent to sewing up a cross-cap \cite{kol}, we consider here the cross-cup in the left panel of Fig.(2) as a spacetime manifold with ${\mathbb{R} P}^2$ topology.

Since we have seen that the spacetime dual to the product state is orientable, the sphere, $S^2$, in this case, it is trivial to determine that decreasing the entanglement to zero is consistent to a topological blow down operation. Roughly speaking, disentangling the degrees of freedom between the two regions implies a transition from a non-orientable spacetime to a spacetime having a definite orientation of space and time, thus an orientable spacetime. We should note here that there is no smooth change of topology of global spacetime, but an abrupt transition from a non-orientable spacetime to an orintable spacetime.

As a final note we ascertain that the continuation past the line of self-intersection is in a region corresponding to the interior of the sphere, a region with reversed time and space. Thus, in the example of eternal black hole, the extended Penrose diagram is interpreted, following the line of the above discussion, as follows: the two asymptotically regions associated to the quadrants I and III correspond, in this case, to the exterior and the interior of the sphere.

\section{\label{sec:level6}Conclusions}

We have discussed the emergence of the arrow of time in the AdS/CFT correspondence framework. We have shown, that the orientation of the thermodynamic arrow of time on dual global spacetimes is dictated by the degree of entanglement on the CFTs side. Accordingly, the lack of initial correlations between the two copies of CFT result in the thermodynamic arrow of time is oriented natural, according to the second law. On the other hand, the initial high-correlation present in the joint state of the CFTs ensures that that the thermodynamic arrow may be oriented in both directions; there is no opportunity for the dominance of one direction of time over the other. Based on these arguments, we have conjectured that the structure of the spacetime geometry is consistent with that of a time-unorionted spacetime, in the presence of the initial high-correlations, while in the case of low initial correlations the geometry is that of a time-oriented spacetime.

Further we relied on the space and time reversal symmetry of the thermofield double state to argue that dual geometric structure of spacetime is  global non-orientable. 

Therefore, starting with the two copies of field theory maximally entangled and disentangling the degrees of freedom is consistent, in the geometrical dual, with an abrupt transition from a non-orientable spacetime to an orientable global spacetime. 

We conclude showing that topology change induced by decreasing the entanglement between two sets of degrees of freedom corresponds to a topological blow down operation.

\end{document}